\documentclass[a4paper,12pt]{article}
\usepackage{amsthm}
\usepackage{amssymb}
\usepackage{amsmath}
\usepackage{color}
\usepackage{amsfonts}

\newtheorem{remark}{Remark}[section]

\newtheorem{lemma}{Lemma}[section]
\newtheorem{theorem}{Theorem}[section]

\newtheorem{corollary}{Corollary}[section]

\def\b1{\mbox{\boldmath $1$}}

\newenvironment{demo*}{\vspace{3mm}\noindent{\bf Proof.}}{\hfill $\Box$ \vspace{3mm}}

\begin{document}

\title{\bf The first passage time problem
 for mixed-exponential
 jump  processes with applications in insurance and finance}
\author{Chuancun Yin\footnote{
Corresponding author. Tel:+865374453221; fax:+865374455076
\newline
 {\it E-mail address:} ccyin@mail.qfnu.edu.cn (C. C. Yin)},\ \ Yuzhen Wen,\ \ Zhaojun Zong,\ \ Ying Shen\\
{\normalsize\it School of Mathematical Sciences, Qufu Normal
University,} {\normalsize\it Shandong 273165,\ China} \\}
\date{ }
\maketitle

 \vskip0.01cm
 \noindent{\large {\bf Abstract}}
  This paper studies the first passage times   to constant  boundaries for
   mixed-exponential jump diffusion processes. Explicit solutions of
  the Laplace transforms of the distribution of the first
  passage times, the joint distribution of the first
  passage times and undershoot (overshoot) are obtained. As applications, we
  present  explicit expression of  the Gerber-Shiu functions for surplus processes with two-sided
  jumps,   present the analytical
solutions for popular path-dependent options such as lookback and
barrier options in terms of Laplace transforms and give a closed-form expression
on the price  of the zero-coupon bond
under a structural credit risk model with jumps.\\



\newpage

%\noindent{\bf 1.~~Introduction}
\section{Introduction}\label{intro}

   One-sided and two-sided exit problems for the compound Poisson processes and
jump diffusion processes with two-sided jumps have been applied
widely in a variety of fields. For example,  in the theory of actuarial
mathematics, the problem of first exit from a half-line is of
fundamental interest with regard to the classical ruin problem and
the expected discounted penalty function  or the Gerber-Shiu
function  as well as the expected total discounted dividends up
to ruin. See e.g. Kl\"uppelberg et al. [1],  Mordecki [2],
Xing et al. [3], Cai et al. [4], Zhang et al. [5], Chi [6] and Chi and Lin [7]. In the setting of mathematical finance, the first passage time plays a crucial role for the pricing of many
path-dependent options, American-type and Russian-type options, see
e.g. Kou [8], Kou and Wang [9, 10], Asmussen et al. [11], Levendorskii [12],
Alili and Kyprianou [13], Cai et al. [14],  Cai and Kou [15]
 as well as certain credit risk models, see, for example,
Hilberink and Rogers [16], Le Courtois and Quittard-Pinon [17] and
Dong et al. [18]. Many optimal stopping strategies also turn out
to boil down to the first passage problem for jump diffusion processes,
see e.g. Mordecki [19].  In queueing theory one-sided and
two-sided first-exit problems for the compound Poisson processes and
jump diffusion processes with two-sided jumps have been playing a
central role in a single-server queueing system with random workload
removal, see e.g. Perry et al. [20]. Usually, when we  study the
first passage problem, the models with two-sided jumps are more
difficult to handle than those with one-sided jumps, because the
undershoot and overshoot problem could not be avoided. Despite the
maturity of this field of study it is surprising to note that, until
very recently, it can only  be solved for certain kinds of jump
distributions, such as the Kou's double exponential jump diffusion
model (see Kou [8], Kou and Wang [9]).
Recently, Cai and Kou [15] proposed a  mixed-exponential jump diffusion process to model the asset return and  found an expression for the  joint distribution of the first passage time and the overshoot for a mixed-exponential jump diffusion process.    In the most recent paper of Wen and Yin [21], two-sided first-exit problem for a jump process having jumps with rational Laplace transform was studied. However,  determination of
 the  coefficients in  expressions of above two papers    still remains a mathematical and computational challenge.
In this paper we will further study the first passage problems in  Cai and Kou [15] and give an explicit expression for the  joint distribution of the first passage time and the overshoot for a mixed-exponential jump process with or without a diffusion.  Moreover, we present several applications in insurance risk theory and in finance.

The rest of the paper is organized as follows.  In Section 2, the model assumptions are formulated. In Section 3, we
study the one-sided  passage  problem from below or above   for compound Poisson
process and jump diffusion process. In Section 4, we give explicit expression of  the Gerber-Shiu function with two-sided jumps. In Section 5, we present the analytical solutions to the pricing problem of one barrier options and lookback options, and in the last section we derive a closed-form expression for the price  of the zero-coupon bond.

  \vskip 0.2cm
\section{Mathematical  model}\label{problem}
\setcounter{equation}{0}

A jump diffusion process $X=\{X(t):t\ge 0\}$ is defined as
\begin{equation}
X(t)=x+\mu t+\sigma W_t+\sum_{i=1}^{N_t}Y_i,
\end{equation}
where $x$ is the starting point of $X$, $\{W_t; t\ge 0\}$ is a standard Brownian motion with $W_0=0$,
$\{N_t; t\ge 0\}$ is a Poisson process with rate $\lambda$,
constants $\mu\in\Bbb{R}, \sigma\ge 0$ represent the drift and the
volatility of the diffusion part respectively, and the jump sizes
$\{Y_i; i\ge 1\}$ are independent and identically distributed random
variables. We assume that  $\{Y_i; i\ge 1\}$ are identically distributed as the canonical r.v. $Y$ with  probability density function  $f_Y(y)$.
Moreover, it is assumed that $\{W_t\}, \{N_t\}$ and $\{Y_i\}$ are
independent. When $\sigma=0$, the process (2.1) is the so called the
compound Poisson process with positive and negative jumps and linear
deterministic decrease or increase between jumps according to
$\mu<0$ or $\mu>0$.
The processes cover many models appearing in the literature such as the compound
 Poisson risk models, the perturbed compound Poisson risk models, and their dual models.
 From now on, we shall denoted by $\{P_x : x\in \Bbb{R}\}$ probabilities such that under $P_x$, $X(0)=x$ with
probability one. Moreover,  $E_x$ will be the expectation operator
associated to $P_x$. For convenience, we shall write $P=P_0$ and
$E=E_0$.

It is easy to see that $X$ is a special case of  L\'evy processes with
two-sided jumps, whose infinitesimal generator of $X$ is given by
\begin{equation}
{\cal{L}}g(x)=\frac12\sigma^2g''(x)+\mu
g'(x)+\lambda\int_{-\infty}^{\infty}(g(x+y)-g(x))f_Y(y)dy,\nonumber
\end{equation}
 for any twice continuously differentiable function $g$. The moment generating function of $X(t)$ is
 $E(e^{z X(t)})=e^{\psi (z) t},\ t\ge 0,  \Re(z)=0$, where
 $\psi(z)$, called the exponent of the L\'evy process $X$, is defined as
 \begin{equation}
 \psi (z)=\frac12\sigma^2 z^2+\mu
z+\lambda(E[e^{z Y}]-1).
\end{equation}
  For more
 about general L\'evy processes, we refer to Bertoin [22],  Kyprianou [23] and
 Doney [24].

\vskip 0.2cm
\section{First passage problems}\label{problem}
\setcounter{equation}{0}

We now turn to one-sided passage problems for the L\'evy
process (2.1). For two flat barriers $h$ and $H$ ($h<H$), define the
first downward passage time under $h$ and the first upward passage
time over $H$ by
$$\tau_h^-:=\inf\{t\ge 0:X(t)\le h\},\; \tau_H^+:=\inf\{t\ge 0:X(t)\ge
H\},$$
with the convention that $\inf \emptyset=\infty$.
In the next two subsections we will investigate the distributions of the
following quantities:  first upward passage time $\tau_H^+$ and
  overshoot $X(\tau_H^+)-H$;
 first downward  passage time $\tau_h^-$ and undershoot  $h-X(\tau_h^-)$.

\subsection{One sided exit from above }

In this subsection we assume that the downward jumps have an
arbitrary distribution with density $f_-$ and Laplace transform $\hat{f}_-$, while the
upward jumps are mixed-exponential, i.e.
\begin{equation}
f_Y(y)=pf_-(-y)\text{\bf
1}_{\{y<0\}}+q\sum_{i=1}^m p_i\eta_i e^{-\eta_i y}\text{\bf
1}_{\{y\ge 0\}},
\end{equation}
 where constants $p, q\ge 0, p+q=1$,   $0<\eta_1<\eta_2<\cdots <\eta_m<\infty$ and $\sum_{i=1}^m p_i=1$.

The L\'evy
exponent of $X$ is given by
$$\psi_1 (z)=\frac12\sigma^2 z^2+\mu
z+\lambda\left(q\sum_{i=1}^m\frac{p_i\eta_i}{\eta_i-z}+p\hat{f}_{-}(-z)-1\right).$$

Using the same argument as in Cai and Kou [15]  we have the following

\begin{lemma}\label{le2-1} (i)\ For sufficiently large $\alpha>0$, if $\sigma>0$ or  $\mu>0$ and $\sigma=0$,  then the equation $\psi_1(z)=\alpha$
 has exactly  $m+1$ distinct positive roots $\beta_1,\cdots,\beta_{m+1}$ satisfying
 $$0<\beta_1<\beta_2<\cdots<\beta_{m+1}<\infty.$$
(ii)\ \ If $\mu\le 0$ and $\sigma=0$, then the equation
$\psi_1(z)=\alpha$
 has  exactly $m$ distinct positive roots  $\beta_1,\cdots,\beta_{m}$
 satisfying
 $$0<\beta_1<\beta_2<\cdots<\beta_m<\infty.$$
\end{lemma}

Cai and Kou [15] found the joint distribution of the first passage time $\tau_H^+$ and  $X(\tau_H^+)$ in the case $\sigma>0$ under the additional assumption $f_{-}(y)$ is also mixed-exponential. However, for a general $f_{-}(y)$ as in the case of the upward jumps are mixed-exponential (cf. Yin, Shen and Wen [25]),
for any sufficiently large $\alpha>0$, $\theta<\eta_1$ and $x<H$, we have
 \begin{eqnarray}
 E_x\left(e^{-\alpha \tau_H^++\theta
X(\tau_H^+)}\right)=\sum_{k=1}^{m+1}w_k e^{\beta_k x},
\end{eqnarray}
where
$w:=(w_1,\cdots,w_{m+1})^{'}$ is a vector uniquely determined by the following system $ABw=J$,
here $A$ is an $(m+1)\times (m+1)$   matrix
$$A=\left[\begin{array}{cccc}1&1&\cdots&1\\
\frac{\eta_1}{\eta_1-\beta_1}&\frac{\eta_1}{\eta_1-\beta_2}&\cdots&\frac{\eta_1}{\eta_1-\beta_{m+1}}\\
\vdots&\vdots&\vdots&\vdots\\
 \frac{\eta_m}{\eta_m-\beta_1}&\frac{\eta_m}{\eta_m-\beta_{2}}&\cdots&\frac{\eta_m}{\eta_m-\beta_{m+1}}
\end{array}
\right],$$
$B$ is an $(m+1)\times (m+1)$ diagonal matrix and $J$ is an $(m+1)$-dimensional vector
$$ B=Diag\{e^{\beta_1 H}, \cdots, e^{\beta_{m+1} H}\},\
 J=e^{\theta H}\left(1, \frac{\eta_1}{\eta_1-\theta},\cdots,\frac{\eta_m}{\eta_m-\theta}\right)^{'}.
$$

In this paper we will determine the coefficients $w_l$'s explicitly. Moreover, we also consider the cases
 $\mu>0, \sigma=0$ and  $\mu\le 0, \sigma=0$.

\begin{theorem} \ For any sufficiently large $\alpha>0$,  we have

(i)\ for $\theta<\eta_1$ and $x<H$,
\begin{eqnarray}
 E_x\left(e^{-\alpha\tau_H^{+}+\theta X(\tau_H^+)}{\bf 1}_{\{\tau_H^+<\infty\}})\right)
=e^{\theta H}\sum_{k=1}^{N}B_k\frac{\prod_{i=1,i\neq k}^{N} (1-\frac{\theta}{\beta_i})}{\prod_{i=1}^{m} (1-\frac{\theta}{\eta_i})}e^{-\beta_k (H-x)},
\end{eqnarray}
(ii)\ for $y\ge 0, x<H$,
 \begin{eqnarray}
 E_x\left(e^{-\alpha\tau_H^+}{\bf 1}_{\{X(\tau_H^+)-H\in dy\}}\right)=
\sum_{k=1}^{N}B_k\left(A_{k0} \delta_0(y)+\sum_{l=1}^{m} A_{kl}\eta_l e^{-\eta_l y}\right)e^{-\beta_k (H-x)}dy,
 \end{eqnarray}
(iii)\ for $x<H$,
  \begin{eqnarray}
  E_x\left(e^{-\alpha\tau_H^+}{\bf 1}_{\{X(\tau_H^+)=H\}}\right)=\sum_{k=1}^{N}B_k A_{k0} e^{-\beta_k (H-x)},
   \end{eqnarray}
(iv)\ for $x<H$, $y\ge 0$,
  \begin{eqnarray}
  E_x\left(e^{-\alpha\tau_H^+}{\bf 1}_{\{X(\tau_H^+)-H>y\}}\right)=\sum_{k=1}^{N}B_k\left(\sum_{l=1}^{m} A_{kl}e^{-\eta_l y}\right)e^{-\beta_k (H-x)},
   \end{eqnarray}
  (v)\ for $x<H$,
  \begin{eqnarray}
  E_x\left(e^{-\alpha\tau_H^+}\right)=\sum_{k=1}^{N}B_k e^{-\beta_k (H-x)},
   \end{eqnarray}
where  $\beta_1,\cdots, \beta_{N}$ are the positive roots of the equation $\psi_1(\beta)=\alpha$, $\delta_0(x)$ is the Dirac delta at $x=0$ and
\begin{equation}
 N=\left\{\begin{array}{lll}&{m}+1,\ \ \ & {\rm if}\ \sigma>0, {\rm or} \ \sigma=0 \ {\rm and}\  \mu>0,\\
&{m},\ \ \ & {\rm if}\ \sigma=0\ {\rm and} \ \mu \le 0,
\end{array}\right.\nonumber
\end{equation}
$$B_j=\frac{\prod_{k=1}^{m} (1-\frac{\beta_j}{\eta_k})}{\prod_{k=1,k\neq j}^{N} (1-\frac{\beta_j}{\beta_k})}, j=1,\cdots, N,$$
\begin{equation}
 A_{k0}=\left\{\begin{array}{lll}& \frac{\prod_{i=1}^{m} \eta_i}{\prod_{i=1, i\neq k}^{N}\beta_i},\  &  {\rm if}\ \sigma>0, {\rm or} \ \sigma=0 \ {\rm and}\  \mu>0,\\
&0,\  &{\rm if}\ \sigma=0\ {\rm and} \ \mu \le 0,
\end{array}\right.\nonumber
\end{equation}
$$A_{kl}=\frac{\prod_{i=1, i\neq k}^{N}(1-\eta_l/\beta_i)}{\prod_{i=1, i\neq l}^{m}(1-\eta_l/\eta_i)},\  l=1,2,\cdots, {m}.$$
\end{theorem}

{\bf Proof} \ We prove the result  for the case $\sigma>0$ only, the rest cases can be proved similarly.  To prove Theorem 3.1, the most difficult part is to find the inverse
  of matrix $A$. For simplicity, we write
$$A=\left[\begin{array}{cc}A_{11}& A_{12}\\
A_{21}& A_{22}
\end{array}
\right],$$
where
$$A_{11}=(1),\; A_{12}=(1,\cdots,1)_{1\times m},\;
A_{21}=\left(\frac{\eta_1}{\eta_1-\beta_1},\cdots,
\frac{\eta_m}{\eta_m-\beta_1}\right)^{'}$$ and
$$A_{22}=\left[\begin{array}{ccc}\frac{\eta_1}{\eta_1-\beta_2}&\cdots&\frac{\eta_1}{\eta_1-\beta_{m+1}}\\
\vdots&\vdots&\vdots\\
\frac{\eta_m}{\eta_m-\beta_{2}}&\cdots&\frac{\eta_m}{\eta_m-\beta_{m+1}}
\end{array}
\right].$$

Note that $A_{22}$ can be written as $A_{22}=J_1 C_1$, where
$J_1=Diag \{\eta_1,\cdots,\eta_m\}$ is a diagonal matrix,
$C_1=\{\frac{1}{\eta_i-\beta_{j+1}}\}_{1\le i,j \le m}$ is a
Cauchy matrix of order $m$ which is invertible   and  the inverse is
given by $C_1^{-1}=\{d_{ij}\}_{m\times m}$, where
$$d_{ij}=(\eta_j-\beta_{i+1})\frac{A_1(\beta_{i+1})}{A_1'(\eta_j)(\beta_{i+1}-\eta_j)}
\frac{B_1(\eta_j)}{B_1'(\beta_{i+1})(\eta_j-\beta_{i+1})}.$$
Here
$$A_1(x)=\prod_{i=1}^m(x-\eta_i),\;\;
B_1(x)=\prod_{i=1}^m(x-\beta_{i+1}).$$ Then the inverse of $A_{22}$
is given by
$$A_{22}^{-1}=\left[\begin{array}{ccc}\frac{1}{\eta_1}d_{11}&\cdots&\frac{1}{\eta_m}d_{1m}\\
\frac{1}{\eta_1}d_{21}&\cdots&\frac{1}{\eta_m}d_{2m}\\
\vdots&\vdots&\vdots\\
\frac{1}{\eta_1}d_{m1}&\cdots&\frac{1}{\eta_m}d_{mm}
\end{array}
\right].$$
The determinant of $C_1$ is given by
(see Calvetti and  Reichel [26]):
$$ \det(C_1)=\frac{\prod_{1\le i<j\le
m}(\eta_i-\eta_j)(\beta_{j+1}-\beta_{i+1})}{\prod_{i,j=1}^m(\eta_i-\beta_{j+1})}.$$
After some algebra,
$$A/A_{22}=
\left(\frac{\prod_{i=1}^m (\beta_{i+1}-\beta_1)}{\prod_{i=1}^m
(\eta_{i}-\beta_1)}\right)_{1\times 1},$$
where
$$A/A_{22}:=A_{11}-A_{12}A_{22}^{-1}A_{21}$$
is the Schur complement of the block $A_{22}$  in $A$, which is a matrix of order 1.
By Schur's  formula (see Zhang [27]),
$$\det(A)=\det(A_{22})\cdot \det(A/A_{22})\neq 0.$$
Moreover, by  Banachiewicz inversion formula (see Zhang [27]), the inverse of $A$ is
given by
 $$A^{-1}=(A/A_{22})^{-1}\left[\begin{array}{cc} 1& -A_{12}A_{22}^{-1}\\
-A_{22}^{-1}A_{21}&A_{22}^{-1}A_{21}
A_{12}A_{22}^{-1}+A_{22}^{-1}(A/A_{22})
\end{array}
\right].$$
After some algebra, we have
$$A_{12}A_{22}^{-1}=\left(\frac{B_1(\eta_1)}{\eta_1 A_1'(\eta_1)}, \cdots,
\frac{B_1(\eta_m)}{\eta_m A_1'(\eta_m)}\right),$$
$$A_{22}^{-1}A_{21}=\left(\sum_{j=1}^m \frac{d_{1j}}{\eta_j-\beta_1}, \cdots,
\sum_{j=1}^m \frac{d_{mj}}{\eta_j-\beta_1} \right)^{'},$$
$$A_{22}^{-1}A_{21}
A_{12}A_{22}^{-1}+A_{22}^{-1}(A/A_{22})
=\left(\frac{B_1(\eta_j)}{\eta_j A_1'(\eta_j)}\sum_{l=1}^m \frac{d_{il}}{\eta_l-\beta_1}+\frac{\prod_{k=1}^m (\beta_{k+1}-\beta_1)}{\eta_j\prod_{u=1}^m
(\eta_{u}-\beta_1)}d_{ij}\right)_{1\le i,j\le  m}.$$
Now  by solving $ABw=J$ we find that
$$w=B^{-1}A^{-1}J=e^{\theta H}\left(B_1\frac{\prod_{i=1,i\neq 1}^{m+1} (1-\frac{\theta}{\beta_i})}{\prod_{i=1}^{m} (1-\frac{\theta}{\eta_i})}e^{-\beta_1 H},\cdots,B_{m+1}\frac{\prod_{i=1,i\neq m+1}^{m+1} (1-\frac{\theta}{\beta_i})}{\prod_{i=1}^{m} (1-\frac{\theta}{\eta_i})}e^{-\beta_{m+1} H}
\right)^{'},$$
from which and (3.2) we get (3.3).

 By the fractional expansion,
  \begin{equation}\frac{\prod_{i=1,i\neq k}^{m+1} (1-\frac{\theta}{\beta_i})}{\prod_{i=1}^{m} (1-\frac{\theta}{\eta_i})}
 =A_{k0}+A_{k1}\frac{\eta_1}{\eta_1-\theta}+\cdots +A_{km}\frac{\eta_m}{\eta_m-\theta},
 \end{equation}
 where the coefficients $A_{kl}$'s are defined in the theorem.
Substituting (3.8) into (3.3) and inverting it on $\theta$ immediately lead to (3.4).  (3.5)-(3.7) are direct consequence of (3.4).
This ends the proof of Theorem 3.1.

{\bf Example 2.1}\  Let $m=1$, several expressions obtained by Theorem 3.1. When  $\sigma>0$ or $\sigma=0$ and
$\mu>0$,  for $x<H, \theta<\eta_1$ and $y\ge 0$, we recover the following
three formulae which obtained by Kou and Wang [10]:
$$E_x\left(e^{-\alpha \tau_H^++\theta
X(\tau_H^+)}\right)=e^{\theta
H}\left(\frac{(\beta_2- \theta)(\eta_1-\beta_1)}{(\eta_1- \theta)(\beta_2-\beta_1)}e^{-\beta_1(H-x)}
+\frac{(\beta_1- \theta)(\beta_2-\eta_1)}{(\eta_1- \theta)(\beta_2-\beta_1)}e^{-\beta_2(H-x)}\right),$$
$$E_x\left(e^{-\delta \tau_H^+}\text{\bf
1}_{\{X(\tau_H^+)-H>y\}}\right) =e^{-\eta_1
y}\frac{(\beta_2-\eta_1)(\eta_1-\beta_1)}
{\eta_1(\beta_2-\beta_1)}\left(e^{-\beta_1 (H-x)}-e^{-\beta_2 (H-x)}\right),$$
$$E_x(e^{-\delta
\tau_H^+})=\frac{\beta_2(\eta_1-\beta_1)}{\eta_1(\beta_2-\beta_1)}e^{-\beta_1(H-x)}
+\frac{\beta_1(\beta_2-\eta_1)}{\eta_1(\beta_2-\beta_1)}e^{-\beta_2(H-x)}.$$
When $\sigma=0$ and $\mu\le 0$,  then for $x<H, \theta<\eta_1$ and
$y\ge 0$,
$$E_x\left(e^{-\delta \tau_H^++ \theta
X(\tau_H^+)}\right)=e^{ \theta
H}\frac{\eta_1-\beta_1}{\eta_1- \theta}e^{-\beta_1(H-x)},$$
$$E_x\left(e^{-\delta \tau_H^+}\text{\bf
1}_{\{X(\tau_H^+)-H>y\}}\right) =e^{-\eta_1
y}\frac{\eta_1-\beta_1}{\eta_1} e^{-\beta_1 (H-x)}.$$

 \subsection{One sided exit from  below }

In this subsection we assume that the upward jumps have an arbitrary
distribution with Laplace transform $\hat{f}_+$, while the downward
jumps are mixed-exponential, i.e.
\begin{equation}
f_Y(y)=p f_+(y)+ q\sum_{j=1}^m p_j\eta_j e^{\eta_j y}\text{\bf
1}_{\{y<0\}},
\end{equation}
 where constants $p, q\ge 0, p+q=1$,   $0<\eta_1<\eta_2<\cdots <\eta_m<\infty$ and $\sum_{j=1}^m p_j=1$.  By (2.2), the
L\'evy exponent of $X$ is given by
$$\psi_2 (z)=\frac12\sigma^2 z^2+\mu
z+\lambda\left(p\hat{f}_+(-z)
+q\sum_{j=1}^m\frac{p_j\eta_j}{\eta_j+z}-1\right).$$

 By replacing $X$ by $-X$ in the previous section, we get the main finding in this section.

\begin{theorem}  \   For any sufficiently large $\alpha>0$,  we have\\
(i)\  for  $\theta>0, x>h$,
\begin{eqnarray}
 E_x\left(e^{-\alpha\tau_{h}^{-}+\theta X(\tau_{h}^-)}{\bf 1}_{\{\tau_{h}^-<\infty\}})\right)
=e^{-\theta h}\sum_{k=1}^{J}B_k\frac{\prod_{i=1,i\neq k}^{J} (1+\frac{\theta}{r_i})}{\prod_{i=1}^{m} (1+\frac{\theta}{\eta_i})}e^{-r_k (x-h)},
\end{eqnarray}
(ii)\ for $x>h, y\ge 0$,
 \begin{eqnarray}
 E\left(e^{-\alpha\tau_{h}^-}{\bf 1}_{\{h-X(\tau_{h}^-)\in dy\}}\right)=
\sum_{k=1}^{J}B_k\left(A_{k0} \delta_0(y)+\sum_{l=1}^{m} A_{kl}\eta_l e^{-\eta_l y}\right)e^{-r_k (x-h)}dy,
 \end{eqnarray}
(iii)\ for $x>h$,
  \begin{eqnarray}
  E_x\left(e^{-\alpha\tau_{h}^-}{\bf 1}_{\{X(\tau_{h}^-)=h\}}\right)=\sum_{k=1}^{J}B_kA_{k0} e^{-r_k (x-h)},
    \end{eqnarray}
(iv)\ for $x>h$,
  \begin{eqnarray}
  E_x\left(e^{-\alpha\tau_{h}^-}{\bf 1}_{\{X(\tau_{h}^-)<h\}}\right)=\sum_{k=1}^{J}B_k\left(\sum_{l=1}^{m} A_{kl}\right)e^{-r_k (x-h)}=\sum_{k=1}^{J}B_k\left(1-A_{k0}\right)e^{-r_k (x-h)},
    \end{eqnarray}
 (v)\ for $x>h$,
  \begin{eqnarray}
  E_x\left(e^{-\alpha\tau_{h}^-}\right)=\sum_{k=1}^{J}B_k e^{-r_k (x-h)},
   \end{eqnarray}
where  $-r_1,\cdots, -r_{J}$ are the negative roots of the equation $\psi_2(r)=\alpha$, and
\begin{equation}
 J=\left\{\begin{array}{lll}&{m}+1,\ \ \ & \sigma>0, {\rm or} \ \sigma=0 \ {\rm and}\  \mu<0,\\
&{m},\ \ \ &\sigma=0\ {\rm and} \ \mu \ge 0,
\end{array}\right.\nonumber
\end{equation}
$$B_j=\frac{\prod_{k=1}^{m} (1-\frac{r_j}{\eta_k})}{\prod_{k=1,k\neq j}^{J} (1-\frac{r_j}{r_k})},\ j=1,\cdots, J,$$
\begin{equation}
 A_{k0}=\left\{\begin{array}{lll}& \frac{\prod_{i=1}^{m} \eta_i}{\prod_{i=1, i\neq k}^{J}r_i},\  & \sigma>0, {\rm or} \ \sigma=0 \ {\rm and}\  \mu>0,\\
&0,\  &\sigma=0\ {\rm and} \ \mu \le 0,
\end{array}\right.\nonumber
\end{equation}
$$A_{kl}=\frac{\prod_{i=1, i\neq k}^{J}(1-\eta_l/r_i)}{\prod_{i=1, i\neq l}^{m}(1-\eta_l/\eta_i)},\  l=1,2,\cdots, {m}.$$
\end{theorem}
\begin{remark}   The result (3.14)  agrees with the  result of Theorem 1.1 in Mordecki [2], where only  the case   $\sigma>0$ and  $p_i\ge 0$ $(i=1,\cdots, {m})$ is considered.
\end{remark}

{\bf Example 2.2}\ Letting $m=1$ in Theorem 3.2. When $\sigma>0$  or
$\sigma=0$ and $\mu<0$,    for $\theta<\eta_1$ and $y\ge 0$,
$$E_x\left(e^{-\alpha \tau_h^-+ \theta
X(\tau_h^-)}\right)=e^{ \theta
h}\left(\frac{(r_2+ \theta)(\eta_1-r_1)}{( \theta+\eta_1)(r_2-r_1)}e^{-r_1(x-h)}
+\frac{(r_1+ \theta)(r_2-\eta_1)}{( \theta+\eta_1)(r_2-r_1)}e^{-r_2(x-h)}\right),
$$
$$E_x\left(e^{-\alpha \tau_h^-}\text{\bf
1}_{\{h-X(\tau_h^-)>y\}}\right) =e^{-\eta_1
l}\frac{(r_2-\eta_1)(\eta_1-r_1)}
{\eta_1(r_2-r_1)}\left(e^{-r_1 (x-h)}-e^{-r_2
(x-h)}\right),$$
$$E_x(e^{-\alpha
\tau_h^-})=\frac{r_2(\eta_1-r_1)}{\eta_1(r_2-r_1)}e^{-r_1(x-h)}
+\frac{r_1(r_2-\eta_1)}{\eta_1(r_2-r_1)}e^{-r_2(x-h)}.
$$
When $\sigma=0$ and $\mu\ge 0$,  then for $ \theta<\eta_1$ and $y\ge 0$,
$$
E_x\left(e^{-\alpha \tau_h^-+ \theta
X(\tau_h^-)}\right)=e^{ \theta
h}\frac{\eta_1-r_1}{ \theta+\eta_1}e^{-r_1(x-h)},
$$
$$E_x\left(e^{-\alpha \tau_h^-}\text{\bf
1}_{\{h-X(\tau_h^-)>y\}}\right) =e^{-\eta_1
y}\frac{\eta_1-r_1}{\eta_1} e^{-r_1 (x-h)}.$$

 \setcounter{equation}{0}
\section{Applications to Gerber-Shiu functions}\label{insu}

We consider an insurance risk model in which the insurer's surplus process is defined as
\begin{equation}
U(t)=u+\mu t+\sigma W_t+\sum_{i=1}^{N_t}Y_i\equiv u+X(t)-x, \ t\ge 0,
\end{equation}
where $X(t)$ is defined by (2.1)   with jump density (3.9). The
time of (ultimate) ruin is defined as $\tau=\inf\{t\ge 0: U(t)\le 0\}$,
where $\tau=\infty$  if ruin does not occur in finite time.
As applications, we  obtain the following special case of   the Gerber-Shiu functions for surplus processes with two-sided jumps.
$$\phi(u)=E(e^{-\alpha \tau}w(|U(\tau)|)1(\tau<\infty)|U(0)=u),$$
$$\phi_d(u)=E(e^{-\alpha \tau}w(|U(\tau)|)1(\tau<\infty,U(\tau)=0)|U(0)=u),$$
$$\phi_s(u)=E(e^{-\alpha \tau}w(|U(\tau)|)1(\tau<\infty, U(\tau)<0)|U(0)=u),$$
where $\alpha>0$ is interpreted as the force of interest and $w$ is a non-negative function defined on $[0,\infty)$. Note that a more general form of Gerber-Shiu function was originally introduced in Gerber and Shiu [28] for the classical risk model.

From Theorem 3.2 (ii) we get the following result.

\begin{corollary}  \   Suppose that $U(t)$ drifts to $+\infty$, then  we have
\begin{equation}
\phi(u)=\int_0^{\infty}w(y)K_u^{(\alpha)}(y)dy,
\end{equation}
 \begin{equation}\phi_d(u)=w(0)\sum_{k=1}^{J}B_kA_{k0}e^{-r_k u},
 \end{equation}
 \begin{equation}\phi_s(u)=\sum_{k=1}^{J}B_k  \left(\sum_{l=1}^{m} A_{kl}\eta_l \int_0^{\infty}w(y) e^{-\eta_l y}dy\right)e^{-r_k u},
 \end{equation}
where $B_k$'s, $A_{kl}$'s and $r_k$'s are defined as in Theorem 3.2, and
$$K_u^{(\alpha)}(y)=\sum_{k=1}^{J}B_k\left(A_{k0} \delta_0(y)+\sum_{l=1}^{m} A_{kl}\eta_l e^{-\eta_l y}\right)e^{-r_k u}.$$
%Here  $r_k$'s are defined as in Theorem 3.2.
\end{corollary}
\begin{remark} We compare our results with the existing literature.  For the case of  $\sigma=0$ and $Y$ has a double exponential distribution,  the result (4.2) was found by Cai et al [4]; For $\sigma=0$ and $\mu=0$,  the result (4.2) was found by Albrecher et al. [29,  (3.2)]; For  $\mu=0$,  the result (4.2) was found by  Albrecher et al. [29,  (9.3)];  For $\sigma=0$ and $\mu<0$,  the results (4.2)-(4.4) were found by Cheung (see Albrecher et al. [29, PP.443-444)].
 \end{remark}

\setcounter{equation}{0}
\section{Applications to pricing path-dependent options}\label{fina}

As applications of our model in finance, we will study the risk-neutral price of
 barrier and lookback options.
These options have a fixed maturity $T$ and a payoff that depends on
the maximum (or minimum) of the asset price on [0, T]. The asset
price process $\{S(t): t\ge 0\}$ under a risk-neutral probability
measure $\Bbb{P}$ is assumed as $S(t)=e^{X(t)}$, where $X(t)$ is
given by (2.1), $S(0)=e^{X(0)}:=S_0$.  We are going to
derive pricing formulae for standard single barrier options and
lookback options, based on the results obtained in Section 3.

\subsection{Lookback options}

The value of a lookback option depends on the maximum or minimum of
the stock price over the entire life span of the option.
 Let the risk-free interest rate be
$r>0$.
 Given a strike
price $K$ and the maturity $T$, it is well-known that (see e.g.
Schoutens [30]) using risk-neutral valuation and after choosing an
equivalent martingale  measure $\Bbb{P}$ the initial (i.e. $t=0$)
price of a fixed-strike lookback put option is given by
$$L^P_{fix}(K,T)=e^{-r T}{\Bbb{E}}\left(\sup_{0\le t\le T}S(t)-K\right)^+;$$
The initial  price of  a fixed-strike lookback  call option is given
by
$$L^C_{fix}(K,T)=e^{-r T}{\Bbb{E}}\left(K-\inf_{0\le t\le T}S(t)\right)^+;$$
The initial  price of  a floating-strike lookback  put option is given
by
$$L^P_{floating}(T)=e^{-r T}{\Bbb{E}}\left(\sup_{0\le t\le T}S(t)-S(T)\right)^+;$$
The initial  price of  a floating-strike lookback  call option is given
by
$$L^C_{floating}(T)=e^{-r T}{\Bbb{E}}\left(S(T)-\inf_{0\le t\le T}S(t)\right)^+.$$

In the standard Black-Scholes setting, closed-form solutions for
lookback options have been derived by Merton [31] and Goldman et
al. [32]. For the double mixed-exponential jump diffusion model,
Cai and Kou [15] derived the Laplace transforms of the lookback
put option price with respect to the maturity $T$, however, the
coefficients do not determinate explicitly.

We shall only consider lookback put options because lookback call
options can be obtained similarly.  For jump diffusion process (2.1) with jump size density (3.1), the condition  $\eta_1> 1$ is imposed to ensure that  the expectation of $e^{-r t}S(t)$ well
defined.

\begin{theorem} For all sufficiently large $\delta>0$, \\
(i)\ for $K\ge S_0$,
$$\int_0^{\infty}e^{-\delta
T}L^P_{fix}(K,T)dT=\frac{S_0}{r+\delta}\sum_{i=1}^{N}\frac{\prod\limits_{l=1}^m\left(1-\frac{\beta_{i,r+\delta}}{\eta_l}\right)}
{\prod\limits^{N}_{{k=1}, k\neq
i}\left(1-\frac{\beta_{i,r+\delta}}{\beta_{k,r+\delta}}\right)}
\frac{1}{\beta_{i,r+\delta}-1}\left(\frac{S_0}{K}\right)^{\beta_{i,r+\delta}-1},$$
$$(ii)\
 \int_0^{\infty}e^{-\delta
T}L^P_{floating}(T)dT=\frac{S_0}{r+\delta}\sum_{i=1}^{N}\frac{\prod\limits_{l=1}^m\left(1-\frac{\beta_{i,r+\delta}}{\eta_l}\right)}
{\prod\limits^{N}_{{k=1}, k\neq
i}\left(1-\frac{\beta_{i,r+\delta}}{\beta_{k,r+\delta}}\right)}
\frac{1}{\beta_{i,r+\delta}-1}+\frac{S_0}{r+\delta}-\frac{S_0}{\delta},$$
where $\beta_{1,r+\delta},\cdots, \beta_{N,r+\delta}$ are the
$N$ positive roots of the equation $\psi_1(z)=r+\delta$, and
\begin{equation}
 N=\left\{\begin{array}{lll}&{m}+1,\ \ \ & \sigma>0, {\rm or} \ \sigma=0 \ {\rm and}\  \mu>0,\\
&{m},\ \ \ &\sigma=0\ {\rm and} \ \mu \le 0.
\end{array}\right.\nonumber
\end{equation}
\end{theorem}

{\bf Proof}\ (i).\ We prove it along the same line as in Cai and Kou [15]. Set $k=\ln\frac{K}{S_0}\ge 0$, then
$$L^P_{fix}(K,T)=S_0 e^{-r T}\int_k^{\infty}e^y\Bbb{P}\left(\sup_{0\le
s\le T}X(s)\ge y\right)dy.$$ It follows that
\begin{equation}
\begin{array}{lll}
\int_0^{\infty}e^{-\delta T}L^P_{fix}(K,T)dT&=&S_0\int_k^{\infty}e^y
\left[\int_0^{\infty}e^{-(r+\delta)T}
\Bbb{P}\left(\sup_{0\le s\le T}X(s)\ge y\right)dT\right]dy\\
&=&\frac{S_0}{r+\delta}\int_k^{\infty}e^y
\Bbb{E}(e^{-(r+\delta)\tau_y^+})dy.\end{array}
\end{equation}
 The
result follows from Theorem 3.1 and (5.1). \\
(ii).\ Since $$L^P_{floating}(T)=S_0 e^{-r
T}{\Bbb{E}}\left[\exp\left(\sup_{0\le t\le
T}X(t)\right)\right]-S_0,$$ it follows that
\begin{equation}
\begin{array}{lll}
\int_0^{\infty}e^{-\delta T}
L^P_{floating}(T)dT&=&S_0\int_0^{\infty}e^{-(r+\delta)T}{\Bbb{E}}\left[\exp\left(\sup_{0\le
t\le T}X(t)\right)\right]dT-\frac{S_0}{\delta}\\
&=&\frac{S_0}{r+\delta}{\Bbb{E}}\left[\exp\left(\sup_{0\le t\le
e(r+\delta)}X(t)\right)\right]-\frac{S_0}{\delta}\\
&=&\frac{S_0}{r+\delta}\left[1+\int_0^{\infty}e^y
\Bbb{P}\left(\sup_{0\le s\le e(r+\delta)}X(s)\ge
y\right)dy\right]-\frac{S_0}{\delta}\\
&=&\frac{S_0}{r+\delta}\left[1+\int_0^{\infty}e^y \Bbb{E}(e^{-(r+\delta)
\tau_y^+})dy\right] -\frac{S_0}{\delta}.
\end{array}
\end{equation}
The result follows from Theorem 3.1 and (5.2).

\subsection{Barrier  options }
The generic term barrier options refers to the class of options
whose payoff depends on whether or not the underlying prices hit a
prespecified barrier during the options' lifetimes. There are eight
types of (one dimensional, single) barrier options: up (down)-and-in
(out) call (put) options. For more details, we refer the reader to
Schoutens [30]. Kou and Wang [10] obtain closed-form price of
up-and-in call barrier option  under a double exponential jump
diffusion model; Cai and Kou [15] obtain closed-form expressions
of the up-and-in call barrier option  under a double
mixed-exponential jump diffusion model. Here, we only illustrate how
to deal with the down-and-out call barrier option  because the other
seven barrier options can be priced similarly. For jump diffusion process (2.1) with jump size density (3.9),  given a strike price
$K$ and a barrier level $U$,  under the risk-neutral probability
measure $\Bbb{P}$, the price of down-and-out call option is defined
as
$$DOC=\exp(-r T){\Bbb{E}}[(S(T)-K)^+ {\bf 1}_{(\inf_{0\le t\le
T}S(t)>U)}|S_0],\ U<S_0.
$$
Let $h=\ln\frac{U}{S_0}$ and $k=-\ln K$. Then
$$DOC(k,T):=DOC=\exp(-r T){\Bbb{E}}_x [(S_0 e^{X(T)}-e^{-k})^+ {\bf
1}_{(\tau^-_h>T)}].
$$
\begin{theorem} For any $0<\phi<\eta_1-1$ and
$r+\varphi>\psi_1(\phi+1)$, then
 $$
\int_0^{\infty}\int_{-\infty}^{\infty}e^{-\phi k-\varphi
T}DOC(k,T)dkdT=\frac{S_0^{\phi+1}\left(1-e^{-(\phi+1)(x-h)}\sum_{k=1}^J B_{r+\varphi,k}e^{-R_k (x-h)}\right)}{\phi(\phi+1)(\varphi+r-\psi_1(\phi+1))}
$$
where
where  $-R_1,\cdots, -R_{J}$ are the negative roots of the equation $\psi_2(r)=r+\varphi$, and
\begin{equation}
 J=\left\{\begin{array}{lll}&{m}+1,\ \ \ & \sigma>0, {\rm or} \ \sigma=0 \ {\rm and}\  \mu<0,\\
&{m},\ \ \ &\sigma=0\ {\rm and} \ \mu \ge 0,
\end{array}\right.\nonumber
\end{equation}
$$B_{r+\varphi,k}=\frac{\prod_{k=1}^{m} (1-\frac{R_j}{\eta_k})}{\prod_{k=1,k\neq j}^{J} (1-\frac{R_j}{R_k})}\cdot\frac{\prod_{i=1,i\neq k}^{J} (1+\frac{\phi+1}{R_i})}{\prod_{i=1}^{m} (1+\frac{\phi+1}{\eta_i})}.$$
\end{theorem}

{\bf Proof}\ Using the same argument as that of the proof of Theorem
5.2 in Cai and Kou [15], we get
\begin{equation}
\begin{array}{lll}
\int_0^{\infty}\int_{-\infty}^{\infty}e^{-\phi k-\varphi
T}DOC(k,T)dkdT&=&\int_0^{\infty}\int_{-\infty}^{\infty}e^{-\phi
k-(r+\varphi)T}{\Bbb{E}}_x [(S_0 e^{X(T)}-e^{-k})^+ {\bf
1}_{(\tau^-_h>T)}]dkdT\\
&=&\frac{S_0^{\phi+1}}{\phi(\phi+1)}\frac{1}{\varphi+r-\psi_1(\phi+1)}
\left(1-{\Bbb{E}}_x[e^{-(r+\varphi)\tau_h^-
+(\phi+1)X(\tau_h^-)}]\right),\nonumber \end{array}\end{equation}
and the result follows from Theorem 3.2(i).

\setcounter{equation}{0}
\section{The price  of the zero-coupon bond}\label{fina}

In this section, we give a simple application on the price  of the zero-coupon bond
under a structural credit risk model with jumps. As in Dong et al. [18], we assume that the total market value of a firm under the pricing probability measure $P$ is given by
$$V(t)=V_0 e^{X(t)-x},\ t\ge 0,$$
where $V_0$ is positive constant,  $X(t)$ is defined as (2.1). For $K>0$, define the default time as
$$\tau=\inf\{t:V(t)\le K\}.$$
 If we set $x=-\ln (K/V_0)$, then
 $$\tau=\inf\{t:X(t)\le 0\}.$$
Given $T>0$ and a short constant rate of interest $r>0$, Dong et al. (2011) shown that the Laplace transform of the fair price $B(0,T)$ of a defaultable zero-coupon bound at time 0 with maturity $T$ is given by
\begin{equation}
\hat{B}(\gamma)=\frac{1-E[e^{-(\gamma+r)\tau]}}{\gamma+r}+\frac{RE[e^{-(\gamma+r)\tau}V(\tau){\bf 1}(\tau<\infty)]}{K\gamma},\nonumber
\end{equation}
where $R\in [0,1]$ is a constant.
When the jump size distribution is a  double hyperexponential distribution, a closed-form expression is obtained,
but the coefficients can not determined explicitly (except for $n=2$). Now  applying the result in Section 3.2, we get the following result:

\begin{corollary}  If  the process $X(t)$ is defined as (2.1) has jump size density (3.9), we have
\begin{eqnarray}
\hat{B}(\gamma)=\frac{1-\sum_{j=1}^{J}C_j e^{-\rho_j x}}{\gamma+r}+\frac{R}{\gamma}\sum_{j=1}^{J}C_j\frac{\prod_{i=1,i\neq j}^{J} (1+\frac{1}{\rho_i})}{\prod_{i=1}^{m} (1+\frac{1}{\eta_i})}e^{-\rho_j x},\nonumber
\end{eqnarray}
where  $-\rho_1,\cdots, -\rho_{J}$ are the negative roots of the equation $\psi_2(\rho)=\gamma+r$, and
\begin{equation}
 J=\left\{\begin{array}{lll}&{m}+1,\ \ \ & \sigma>0, {\rm or} \ \sigma=0 \ {\rm and}\  \mu<0,\\
&{m},\ \ \ &\sigma=0\ {\rm and} \ \mu \ge 0,
\end{array}\right.\nonumber
\end{equation}
$$C_j=\frac{\prod_{k=1}^{m} (1-\frac{\rho_j}{\eta_k})}{\prod_{k=1,k\neq j}^{J} (1-\frac{\rho_j}{r_k})},\ j=1,\cdots, J.$$
\end{corollary}

\setcounter{equation}{0}
\section{Conflict of Interests}

The authors declare that there is no conflict of interests
regarding the publication of this paper.

\vskip0.3cm

\noindent{\bf Acknowledgements}\; The authors are grateful to the anonymous referee's careful
reading and detailed helpful comments and constructive suggestions,
which have led to a significant improvement of the
paper.  The research  was supported by
the National Natural Science Foundation of China (No. 11171179) and
the Research Fund for the Doctoral Program of Higher Education of
China (No. 20133705110002).

\end{document}